\documentclass[
reprint,
superscriptaddress,
prl,
amsmath,amssymb,
aps,
]{revtex4-2}
\usepackage[T1]{fontenc}     
\usepackage{graphicx}
\usepackage[caption=false]{subfig}
\usepackage{bm}
\usepackage{color}
\usepackage{tikz}

\usepackage{filecontents}

\makeatletter
\@input{SIAUX.tex}
\makeatother

\usepackage{xr-hyper}

\begin{document}
	
	\preprint{APS/123-QED}
	
	\title{Controlling Topological Defects  in Polar Fluids via Reinforcement Learning}
	
	\author{Abhinav Singh}
	\email{abhinav@seas.harvard.edu}
	\affiliation{School of Engineering and Applied Sciences, Harvard University, Cambridge, MA, USA}
	\author{Petros Koumoutsakos}%
	\affiliation{School of Engineering and Applied Sciences, Harvard University, Cambridge, MA, USA}
	\makeatletter
	\newcommand*{\rom}[1]{\expandafter\@slowromancap\romannumeral #1@}
	\makeatother
	
	\begin{abstract}   

Topological defects in active polar fluids exhibit complex dynamics driven by internally generated stresses, reflecting the deep interplay between topology, flow, and non-equilibrium hydrodynamics. Feedback control offers a powerful means to guide such systems, enabling transitions between dynamic states. We investigated closed-loop steering of integer-charged defects in a confined active fluid by modulating the spatial profile of activity. Using a continuum hydrodynamic model, we show that localized control of active stress induces flow fields that can reposition and direct defects along prescribed trajectories by exploiting non-linear couplings in the system. A reinforcement learning framework is used to discover effective control strategies that produce robust defect transport across both trained and novel trajectories. The results highlight how AI agents can learn the underlying dynamics and spatially structure activity to manipulate topological excitations, offering insights into the controllability of active matter and the design of adaptive, self-organized materials.
	\end{abstract}
    
	\maketitle
	
	\begin{figure}
		\centering
		\begin{tikzpicture}[
			neuron/.style={circle, draw, minimum size=0.3cm},
			layer/.style={rectangle, draw=none, minimum height=0cm, minimum width=0.0cm},
			]
			\node (f1a) at (-5,0) {\includegraphics[width=2.5cm]{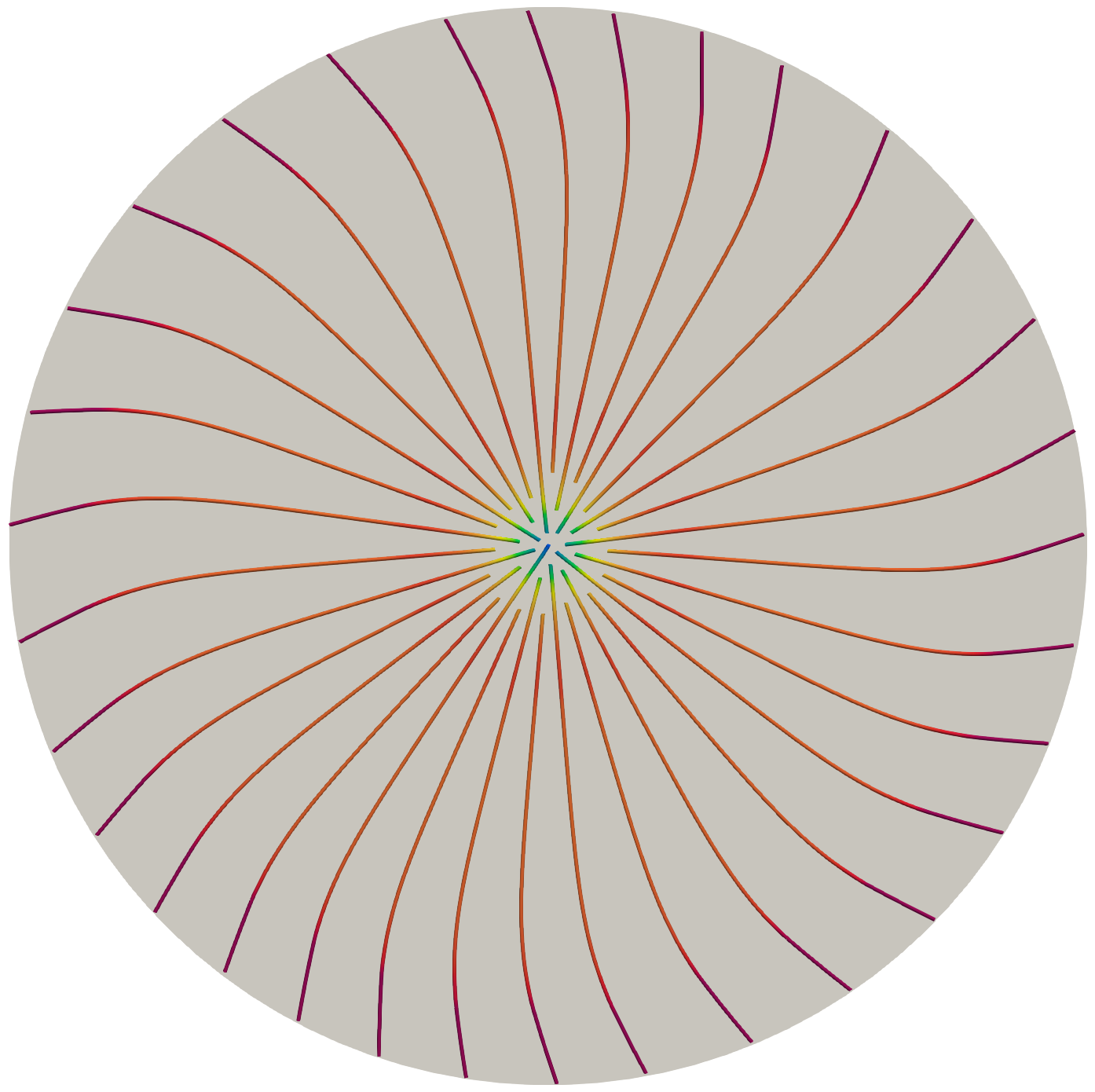}};
			
			\node[layer] (L1) at (-3.25,1) {};
			\node[layer] (L2) at (-2.25,1) {};
			\node[layer] (L3) at (-1.25,1) {};
			
			\foreach \i in {1,2,3} {
				\node[neuron] (L1\i) at (L1 |- 0,1.5-0.5*\i+0.5) {};
				\node[neuron] (L2\i) at (L2 |- 0,1.5-0.5*\i+0.5) {};
				\node[neuron] (L3\i) at (L3 |- 0,1.5-0.5*\i+0.5) {};
			}
			\foreach \i in {1,2,3} {
				\foreach \j in {1,2,3} {
					\draw[-] (L1\i) -- (L2\j);
					\draw[-] (L2\i) -- (L3\j);
				}
			}
			
			\node (f1b) at (0.5,0) {\includegraphics[width=2.5cm]{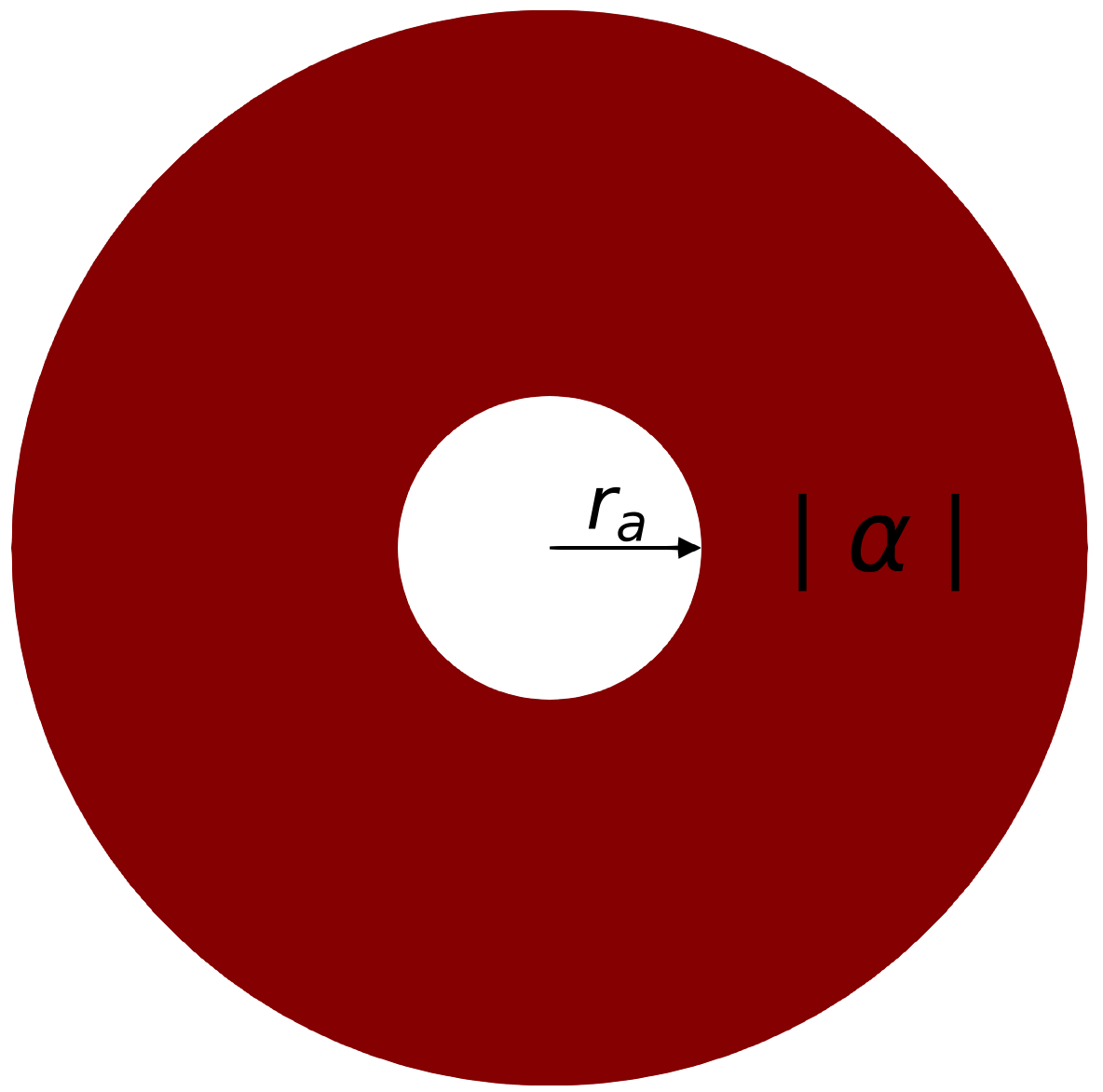}};
			
			\node (f1c) at (-2.25,-1.25) {\includegraphics[width=2.5cm]{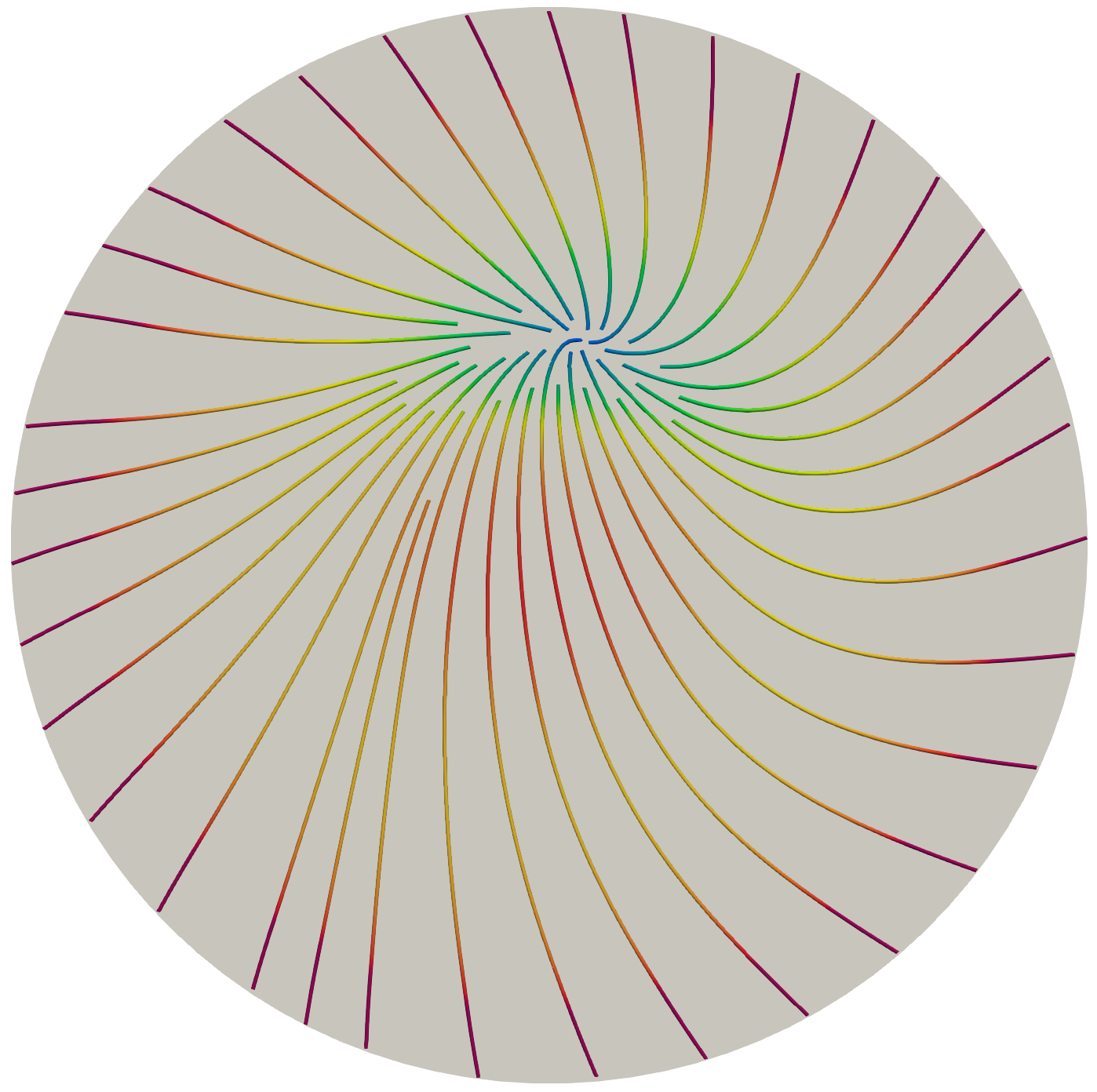}};
			
			\draw[->, bend left=30] ([xshift=1cm,yshift=-0.5cm]f1a.north) to (L12);
			\draw[->, bend left=30 ] ([xshift=-1cm,yshift=0.5cm]f1b.south) to (f1c.east);
			\draw[->, bend left=30] (L32.east) to ([xshift=-1cm,yshift=-0.5cm]f1b.north);
			\draw[<-, bend left=30] (f1c.west) to ([xshift=1cm,yshift=0.5cm]f1a.south);
			
			\node[below] at (f1a.south) {State at $t$};
			\node[below] at (f1b.south) {Spatiotemporal};
			\node[below] at ([yshift=-0.35cm]f1b.south) {activity $\alpha(r,t)$};
			\node[below] at (f1c.south) {State at $t+\delta t$};
		\end{tikzpicture}
		\caption{Spatiotemporal control of a topological defect via RL controlling activity strength and activation region. The location of the defect and its instantaneous velocity are input to a neural network that describes a spatiotemporal activity field as an action leading to closed loop control of the defect position.}
		\label{fig:rl}
	\end{figure}
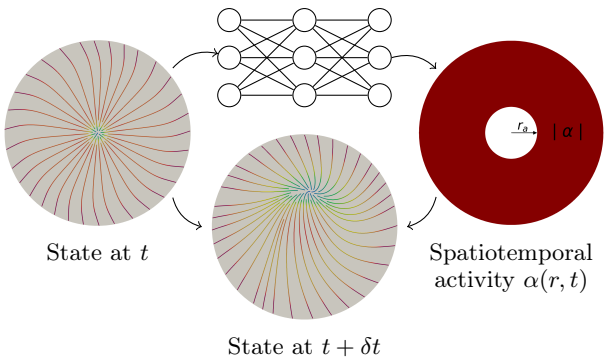
	   
		\begin{figure*}
		\includegraphics[scale=0.15]{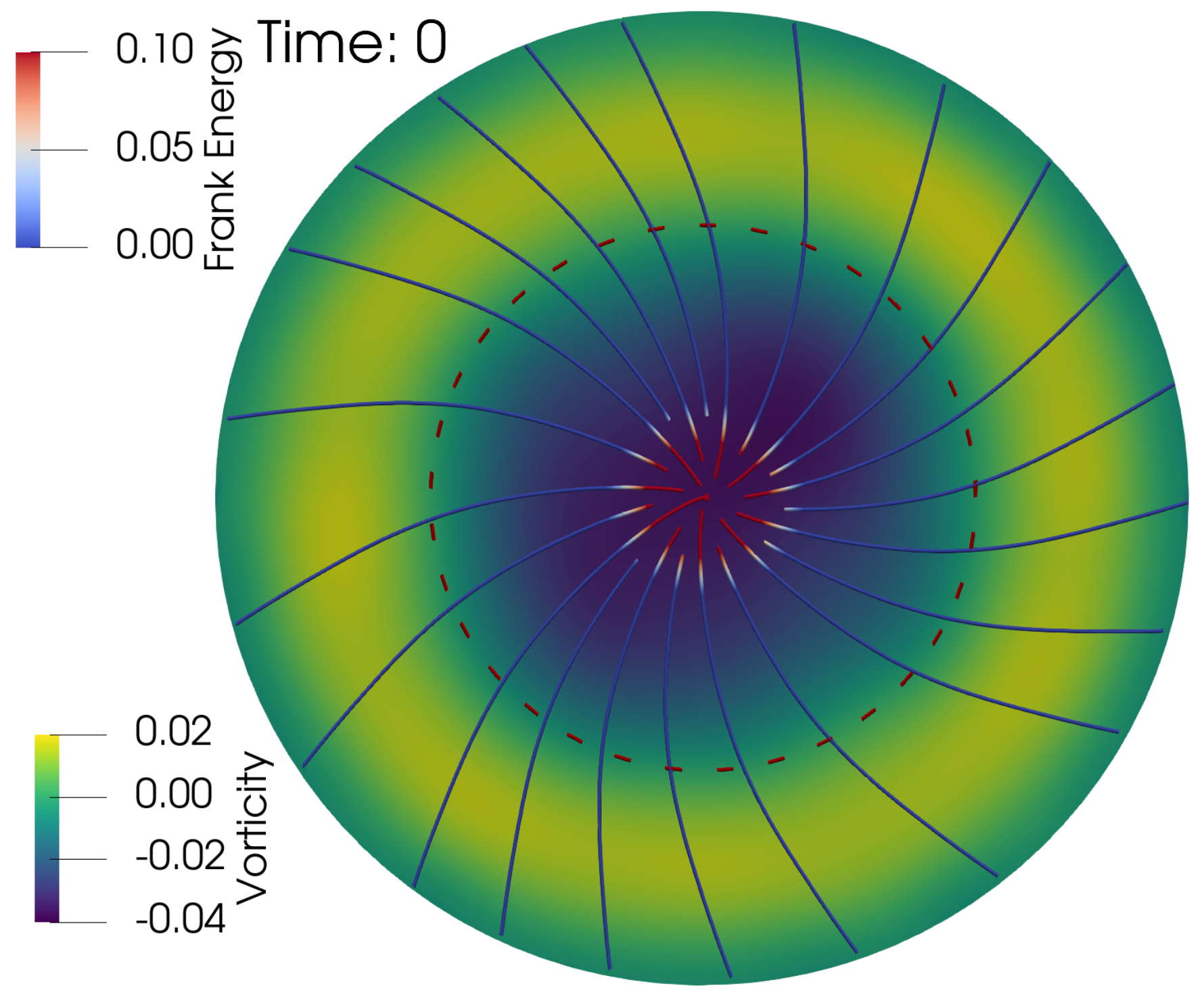}
		\includegraphics[scale=0.15]{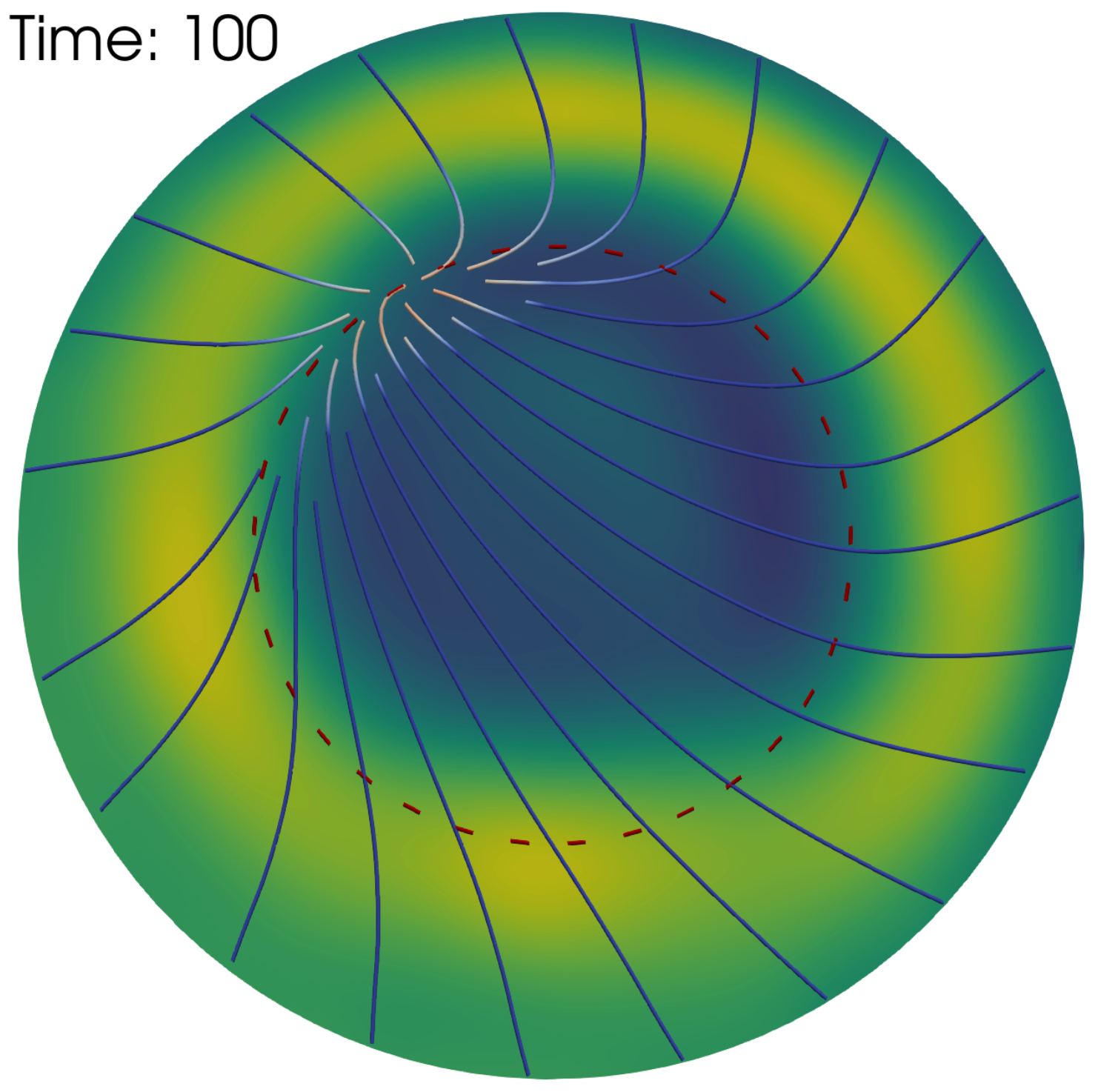}
		\includegraphics[scale=0.15]{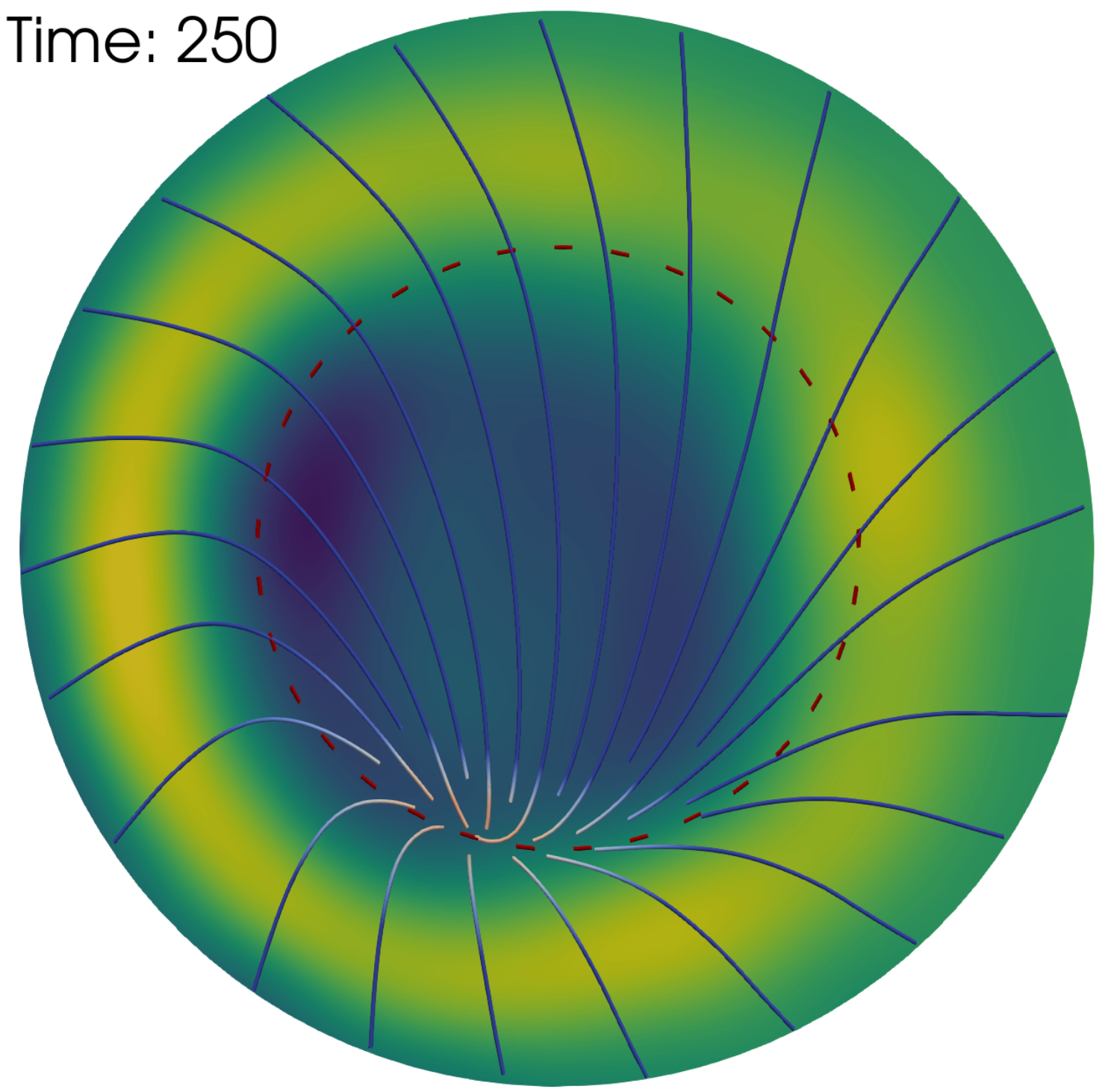}
		\includegraphics[scale=0.15]{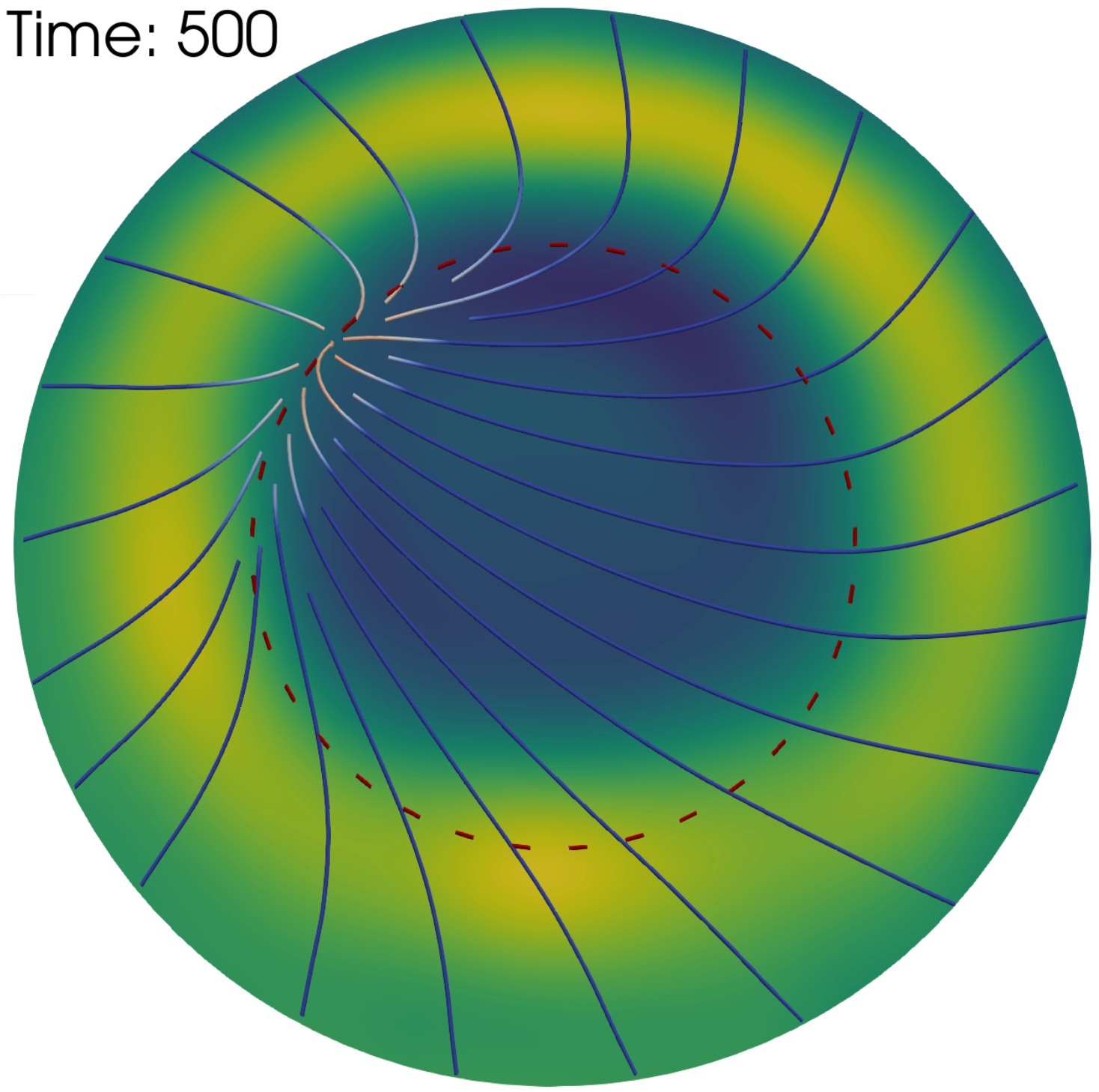}
		\caption{Spontaneous rotation of a +1 charged active polar defect. For a fixed activation radius $r_a$, and a given activity strength $\alpha$, the defect moves from the center of the disk to a fixed radial location $r_d$, and rotates with angular velocity $\omega$. The defect trajectory is shown as a red dashed line. The polarity is shown with streamlines colored with Frank energy and the background is colored with vorticity.}
		\label{fig:f2}
	\end{figure*}
	
Active matter encompasses a diverse class of systems and materials composed of interacting, energy-consuming units. Examples range from cytoskeletal fibers and tissues to flocks of birds and schools of fish, all of which show collective behavior on scales larger than their individual components \cite{Ramaswamy2010,marchetti_hydrodynamics_2013,julicher_hydrodynamic_2018,Saintillan2013}. Through the unique ability to convert chemical energy into mechanical work at the level of individual molecules, constituents, or agents, active matter systems perform complex functions that are often regulated by feedback control. Biological systems provide canonical examples: Living bacteria exhibit chemotaxis, a biophysical feedback mechanism that biases their motion toward nutrient sources, which, on a collective scale, leads to swarming behavior directed by chemical gradients \cite{MaroudasSacks2020}. Other forms of environmental responsiveness include gravitaxis, aerotaxis, magnetotaxis, and phototaxis, each triggering spatiotemporal reorganization in response to external perturbations \cite{guillamat_integer_2022}. In particular, active nematic and polar fluids have attracted significant interest due to their rich defect dynamics and the potential for emergent self-organization \cite{kruse_asters_2004,shankar_hydrodynamics_2019,bowick_symmetry_2021,shankar_topological_2022}. Topological defects in these systems govern mesoscale phenomena and have been implicated in biological processes such as morphogenesis in Hydra and coordinated migration in cell tissues. While these findings underscore the functional potential of defects in active systems, rationally programming or controlling their behavior remains an open challenge. Here, we address this challenge by leveraging reinforcement learning (RL) to guide and manipulate an active polar defect confined within a circular domain.
	
Classical control strategies for active fluids, include boundary confinement \cite{Alam2024,Singh2023-2,Hardoin2022,blanch-mercader_quantifying_2021,Wu2017}, friction anisotropy \cite{Kralj2024,doostmohammadi_coherent_2019}, and optogenetic or light-activated motor complexes \cite{Lemma2023, Ross2019}. They  have provided valuable insight into controlling the trajectories and configurations of defects. Recent studies have focused specifically on steering half-integer charge defects in active nematics with design patterns \cite{Shankar2024} or optimal control \cite{norton_optimal_2020,Ghosh2024,Ghosh2024-2}. More work has been done on understanding energy transfer cascades associated with active length scales \cite{pearce2024}, and employing proportional-integral control for active nematic fluid velocities \cite{Nishiyama2024}. However, these approaches rely predominantly on model predictive control frameworks, which often fail to dynamically reconfigure complex active systems due to slower model-based computations that lack the spatio-temporal precision required for rapid feedback \cite{Takatori2025}. More recently, approaches such as reinforcement learning (RL), have shown remarkable success in control problems in collective hydrodynamics, robotics, navigation in complex flows, and path planning for active Brownian particles \cite{gazzola2016learning,novati_controlled_2019, colabrese2017flow, Novati2021,Gunnarson2021, Karnakov2025,amoudrouz2024}. RL does not require explicit knowledge of the underlying equations of motion, discovers actuation protocols beyond linear regimes, and incorporates closed-loop feedback by continuously adapting to instantaneous system states, such as defect position, velocity, and local-order parameter fields. RL-based strategies present a robust alternative to classical  methods for extending autonomous and self-tuning control to active polar fluid defect trajectories.
	
In this paper we demonstrate the capabilities of RL for the control of active polar fluids. The RL agent receives input about the state of the system (location and velocity of the defect) and outputs a spatiotemporal activity field that modifies the local activity strength $\alpha(r,t)$ (see Fig.~\ref{fig:rl}). This effectively allows us to steer the defect or alter its speed and direction of motion. The RL agent senses the current position of the defect and consults a neural network to  modulate the strength and location of the activity. Over multiple episodes, the RL algorithm adjusts its policy to maximize a cumulative reward, which can be a measure of various properties such as the accuracy, stability, and energy efficiency of defect positioning.
	
We focus on controlling the trajectory of an integer-charged active polar defect confined within a circular domain, subject to no-slip for flow and anchoring conditions for the polar order parameter. Inspired by experimental microtubule motor protein systems that generate active stresses through ATP consumption, utilizing a symmetry-preserving hydrodynamic theory \cite{kruse_asters_2004} to model defect dynamics. Under uniform activity, the defect rests at the center of the circular domain without flow in the system. However, activity gradients generate spontaneous flows that drive the defect to an off-center location, towards higher activity regions. The defect then starts to rotate in the domain as dictated by the rotating boundary conditions. This produces a dynamic circular trajectory of the defect for a given activity gradient $\alpha(r,t)$, as shown in Fig.~\ref{fig:f2} and SI-Movie \rom{1}. 
	
\begin{figure}
		\centering
		\setlength{\tabcolsep}{10pt}
		\captionsetup[subfloat]{captionskip=0pt} 
		\subfloat[]{\includegraphics[scale=0.3]{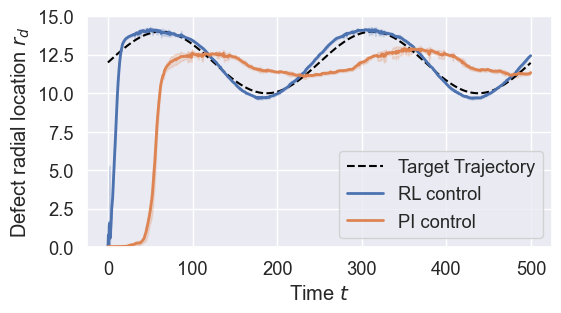}} \subfloat[]{ \includegraphics[scale=0.3]{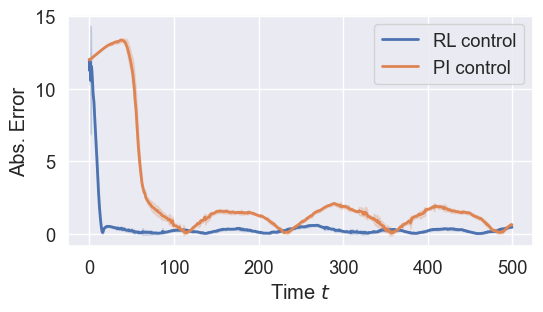}} \\[-3ex]
		\subfloat[]{\includegraphics[scale=0.3]{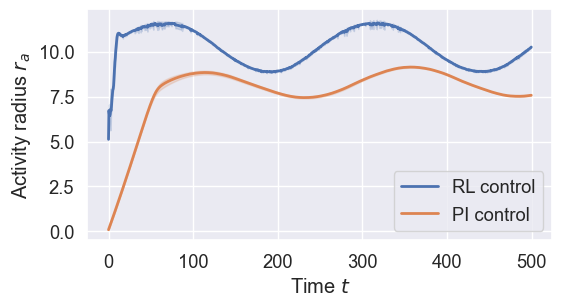}} 	\subfloat[]{\includegraphics[scale=0.3]{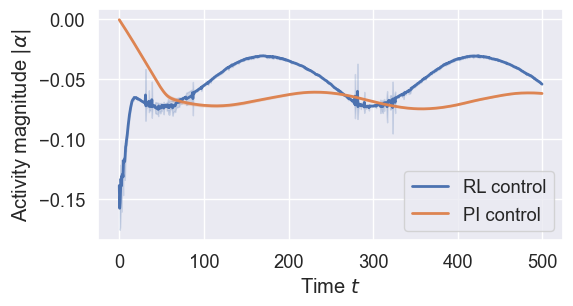}}
		\vspace*{-0.25cm}     
		\caption{ 
			Comparison of RL vs PI control when the RL agent is trained on the target trajectory. Mean and standard deviations over 10 trajectories are shown. (a) Time evolution of the defect’s radial position $r_d$. The dashed line indicates the target trajectory, while the solid lines show the RL-controlled (blue) and PI-controlled (orange) defect positions. (b) Absolute tracking error for the RL (blue) and PI (orange) controllers. The RL approach achieves smaller deviations over time when compared to PI. (c) Time evolution of the activity radius $r_a$ as determined by each controller. (d) Time evolution of the activity strength $\alpha$ as determined by the each controller.}
		\label{fig:train}      
	\end{figure}

To control the defect trajectory, a reinforcement learning agent (RL) is implemented within a simulation environment that emulates the polar active hydrodynamic model. In the present framework, actions are specified by spatiotemporal modulations of the activity field (i.e., activation radii and strength) and are imposed on the system. The environment returns the instantaneous state (e.g., defect position and velocity). The inherent Markovian nature of the simulation, where the current state fully determines the future evolution, makes the defect control problem suited to deep RL \cite{sutton_reinforcement_2020}. The neural network–based agent is trained to maximize a reward function that minimizes the deviation of the defect from a prescribed target trajectory. Notably, while a conventional proportional-integral controller performs adequately for static targets (see SI section \ref{SI-starget}), it is outperformed in tracking dynamic trajectories as shown in Fig.~\ref{fig:train} and SI-Movie \rom{2}-\rom{3}.
	
\begin{figure}
		\centering
		\setlength{\tabcolsep}{10pt}
		\captionsetup[subfloat]{captionskip=0pt} 
		\subfloat[]{\includegraphics[scale=0.3]{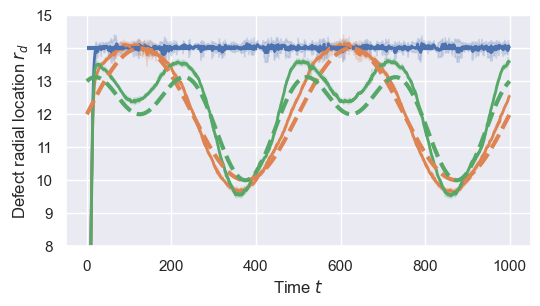}} \subfloat[]{ \includegraphics[scale=0.3]{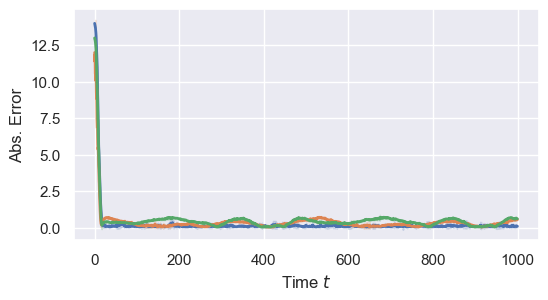}} \\[-3ex]
		\subfloat[]{\includegraphics[scale=0.3]{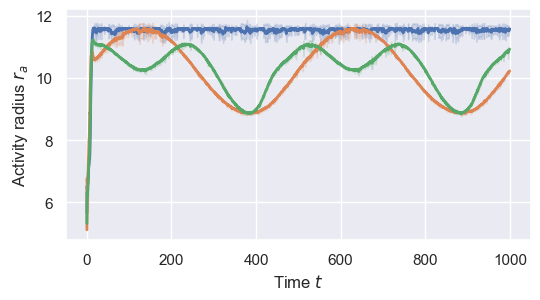}} 	\subfloat[]{\includegraphics[scale=0.3]{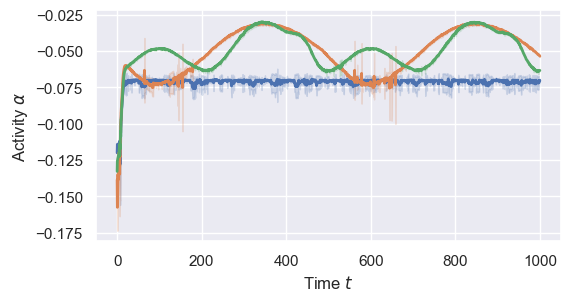}}
		\vspace*{-0.25cm}     
		\caption{Generalization of the RL agent to unseen target trajectories and longer time span $t_f=1000$: $\hat{r}(t)=14$ in blue, $\hat{r}(t)=12+2\sin(4\pi t/t_f)$ in orange, and $\hat{r}(t)=12+\sin(4\pi t/t_f)+\cos(8\pi t/t_f)$. Mean and standard deviations over 10 trajectories are shown. (a) Time evolution of the defect’s radial position $r_d$. The dashed line indicates the corresponding target trajectory, while the solid lines show the RL-controlled defect positions. (b) Absolute tracking error for the RL controller.  (c) Time evolution of the activity radius $r_a$ for each case. (d) Time evolution of the activity strength $\alpha$ as determined by the controller for each case.}
		\label{fig:test}       
	\end{figure}

	Using a state space that includes both the target trajectory and its instantaneous rate of change, equips the RL agent with predictive capacity that goes beyond reactive control. This representation enables the agent to anticipate the future dynamics of the system and generalize its control policy to arbitrary paths, even those not encountered during training. In particular, while the agent was trained on sinusoidal trajectories, this comprehensive state feedback allowed it to robustly track and stabilize the defect along a wide variety of motion patterns. This level of generalization, which contrasts sharply with the limitations of conventional controllers constrained to static or linear regimes, highlights the advantage of this approach, as shown in Fig.~\ref{fig:test} and SI-Movie \rom{4}.
	
	The results show that reinforcement learning can autonomously derive robust control policies to accurately reposition and stabilize active polar defects, despite the complex nonlinear dynamics of active fluids. This RL-based strategy not only complements earlier approaches such as boundary confinement, friction, and manual light-activated manipulations but also offers a versatile platform for achieving adaptive closed-loop control in active systems. The generality of this approach suggests that it could be extended to multidefect environments or arrays of active domains, thereby paving the way for next-generation self-organizing materials with on-demand structural reconfiguration or controlling biophysical phenomena such as morphogenesis.
	
	\textit{Active Polar Hydrodynamic Model}.
	The simulation environment is described by the following active polar hydrodynamic model that describes the time evolution of the polar order parameter, referred to as the polarity vector $\mathbf{p}$,
	\begin{align}
		\frac{\partial \mathbf{p}}{\partial t}+(\mathbf{u}\cdot\nabla)\mathbf{p}+\mathbf{W}\cdot\mathbf{p}=\frac{\mathbf{h}}{\gamma}+\lambda\alpha\mathbf{p}-\nu \mathbf{E}\cdot\mathbf{p}
	\end{align}
	where $\mathbf{E}$, $\mathbf{W}$  are the strain-rate and spin tensors of velocity $\mathbf{u}$ respectively. $\mathbf{h}$ is the molecular field obtained via the variational derivative of the Frank free energy density,
	\begin{align}
		\mathcal{F}=\frac{K_s}{2}(\nabla\cdot\mathbf{p})^2+\frac{K_b}{2}(\nabla\times\mathbf{p})^2+\frac{\chi}{2}|\mathbf{p}|^2(1-\frac{1}{2}|\mathbf{p}|^2)
	\end{align}
	with $K_s$ and $K_b$ as material constants of splay and bend respectively, with $\chi$ to enforce unit magnitude of polarity as a soft-constraint in the system. The force balance that dictates the flow velocity $\mathbf{u}$, is described by
	\begin{align}
		\nabla.(\mathbf{\Sigma^{\mathrm{a}}}+\mathbf{\Sigma^{\mathrm{p}}})=0 && \nabla.\mathbf{u}=0
	\end{align}
	where $\mathbf{\Sigma^{\mathrm{a}}}=\alpha(\mathbf{x},t)(\mathbf{p}^2-\frac{1}{2}\mathbf{I})$ is the active stress driving spontaneous flows and $\mathbf{\Sigma^{\mathrm{p}}}$ is the passive nematic liquid crystal stress comprising of symmetric, anti-symmetric, and equilibrium stress. See SI section \ref{SI-simdetails} for more details.
	$\alpha(\mathbf{x},t)$ is the spatiotemporal chemical potential field that drive hydrodynamic flows.
	The sign of $\alpha(\mathbf{x},t)$ describes type of active stress to be contractile or extensile, and its magnitude describes the strength of the acting forces. Consider a disk with an activity pattern parameterized by $r_a$ as shown in Fig.~\ref{fig:rl}. Hence the activity field is fully described by two parameters representing the cutoff radius $r_a$ and magnitude of the activity, denoted by $\alpha$,
	\begin{equation}
		\alpha(r,t)= 
		\begin{cases} 
			\alpha, & r \geq r_a\\
			0, & r < r_a 
		\end{cases}
	\end{equation}  	
	The nondimensionalization of the equations can be performed by rescaling with the \emph{active length scale} $L_a = \sqrt{\frac{\delta K}{\alpha}}$
and \emph{active time scale} $ \tau_a = \frac{1}{\alpha}$, where $\delta K=K_b-K_s$. Dimensionless variables are described in the SI section \ref{SI-nd}. Choosing the parameters $\eta=1, \nu=2, \gamma=0.5, \lambda=15, K_s=5, K_b=7.5,$ and $\chi=-0.5$ for the simulation environment. The simulation domain is a disk of radius $R=25$. $r_a, \alpha$ is chosen by the agent within a range of $[0,R], [-0.2,0]$ respectively. Active hydrodynamic equations are solved using the OpenFPM opensource library \cite{incardona_openfpm_2019,singh_c_2021}, with a previously described numerical algorithm that discretizes the continuous variables in point clouds and uses a direct numerical simulation approach for the solution \cite{Singh2023}.	
	
	\captionsetup{belowskip=-5pt}
	\textit{An Off-Policy Actor-Critic Approach: VRACER.} A continuous off-policy actor-critic reinforcement learning scheme \cite{Degris2012} is implemented to achieve precise spatiotemporal control over active nematic defects. Here, the control policy is iteratively refined through replay of experiences, where actions are selected to maximize a reward function based on a repository of past interactions. Specifically, the VRACER algorithm \cite{pmlr-v97-novati19a} is applied, which ensures the derivation of stable control policies by carefully updating the policy through the rejection sampling of experiences that fall outside a trust region of possible policy parameters. This strategy effectively maintains stability while promoting robust learning in various control tasks.
	
	The agent’s state space is chosen as $s = \{r_d,v_d,r_a,\alpha,\hat{r}_t,\dot{\hat{r}}_t\}$, which captures the position of the defect, the velocity of the defect, the radius of the activity cut-off, the strength of the activity, the target trajectory and the target trajectory gradient, respectively. In each state, a neural network outputs parameters $\{m^{w}(s),\sigma^{w}(s),V^{w}(s),l^{w}(s)\}$ that define both a Gaussian policy with mean $m^{w}(s)$ and variance $\sigma^{w}(s)$ parametrized by the weights $w$ of the neural network, and the value function of the state $V^{w}(s)$ representing the expected cumulative reward. Here, $l^{w}(s)$ denotes the rate at which the action value decreases for actions farther away from the mean, a quantity that is instrumental for computing importance sampling ratios and ensuring stable, trust region-based updates during off-policy learning. However, we assume that $Q^w(s, a)$ (the state action value function) is maximal at the mean of the policy for every given state.
	
	The policy is defined as
\begin{align}
	\pi_{w}(a|s)=\frac{1}{\sqrt{2\pi}\sigma^{w}(s)}\exp\left[-\frac{(a - m^{w}(s))^{2}}{2[\sigma^{w}(s)]^{2}}\right],
\end{align}	

	and the action-value estimate is
\begin{align}
	Q^{w}(s,a)=V^{w}(s)-\tfrac{1}{2}[l^{w}(s)]^{2}\{(a-m^{w}(s))^{2}-[\sigma^{w}(s)]^{2}\}.
\end{align}
	By continually sampling trajectories and updating the network weights using gradient-based methods \cite{Degris2012}, the agent learns to adapt the action distribution to improve the alignment and stability of the defect. This framework allows efficient exploration of the state-action space in active polar fluids, leveraging data-driven control strategies to navigate complex nonlinear dynamics of active polar defects.  Although the current implementation relies on simulated experiences, this approach could seamlessly extend to continuously acquired experimental data, enabling robust policy learning without the need for explicit parameter fitting of the active hydrodynamics.
	
	To guide the agent toward a desired defect trajectory, instantaneous reward $R_t$ is defined as the distance between the location of the defect and its target location at time t. If the defect position in time step $t$ is  $r_{d}(t)$  and the target position is $\hat{r}_d(t)$,
	\begin{align}
		R_{t} = |r_{d}(t)-\hat{r}_d(t)|
		\label{eq:reward}
	\end{align}
	This choice ensures the agent is rewarded whenever it reduces the deviation from the target. A unique feature of the approach is also self-perception of actions and the target by the agent. This trains the agent to generalize over arbitrary targets that are input to the cotrol policy during inference.
	Additional terms, such as a small penalty proportional to the energy injection input $\int_\Omega \alpha^2 \mathrm{d}^2\mathbf{x}$, can be included to reflect energy costs. Thus, the RL agent learns control policies that minimize the distance between the actual and desired defect trajectories without explicitly enforcing boundary conditions or solving complex optimal control formulations that require detailed model formulation.
	
	Training is carried out over a 1000 episode rollout, where each episode consists of a trajectory from $t=0$ to $t_f=500$ with a time step $\delta t=2.5$.  At each time step, the states are updated based on the current action and the reward is computed as per Eq.~\eqref{eq:reward} for a given training trajectory. The agent is represented by a fully connected feedforward neural network with a single hidden layer comprising 16 units. The active hydrodynamic simulation environment is coupled to the Korali framework \cite{Martin2022}, which is designed for large-scale reinforcement learning. In particular, Korali’s distributed engine is used to run multiple instances of the environment in parallel, where each environment also runs a simulation in parallel using OpenFPM for faster training. For more details, see SI section \ref{SI-rldetails}.

	\textit{Results:}
	First, we examine the ability of the RL agent to maintain the defect at a fixed radial location by prescribing a static target trajectory, $\hat{r}_d(t)=r_0$. In this scenario, the activity magnitude is fixed $|\alpha|=0.05$ and the agent is allowed to adjust the cutoff radius only $r_a$. Physically, the active polar defect is highly sensitive to gradients in the activity field. The cutoff radius $r_a$ determines the spatial extent to which active stresses are applied, and even small changes in $r_a$ can lead to significant changes in the local hydrodynamic forces acting on the defect. By modulating $r_a$, the RL agent effectively sculpts the active stress profile within the circular domain, creating a local flow field that can rapidly push or pull the defect toward the prescribed target, as shown in SI-Fig.~\ref{SI-fig:static}. In contrast, a traditional proportional-integral (PI) controller relies on a fixed feedback structure that integrates the error over time. Although integration helps to eliminate steady-state errors, it also introduces a lag in response. This delay is problematic in active matter systems, where the dynamics are strongly nonlinear and highly sensitive to even minor perturbations. 
	
	The analysis is then extended to a dynamic scenario by training the agent on a rotating sinusoidal trajectory defined as $\hat{r}(t)=r+A\sin(\omega t/t_f)$ with parameters $r=12, A=2, \omega=4\pi$ and $t_f=500$. In this case, the defect must follow a continuously changing target trajectory, which places an even greater demand on the control strategy. The dynamic behavior of active fluids is inherently non-linear: local active stresses generate flows that not only depend on the current state but also evolve rapidly over time. This means that timing is crucial; any delay in adjusting the control parameters can result in significant deviations from the target.
	
	The RL agent addresses this challenge using a predictive control strategy. During training, the agent observes not only the instantaneous position and velocity of the defect but also the target trajectory and its rate of change. This comprehensive state feedback enables the agent to learn an internal model of the dynamics of the defect. When the target trajectory begins to change, the RL agent is capable of preemptively modulating both the cutoff radius $r_a$ and the activity strength $\alpha$ to generate the appropriate spatiotemporal gradients in the active stress field. As a result, the defect is steered along the sinusoidal path with high fidelity. The improved performance is clearly seen in Fig.~\ref{fig:train}a, where the mean trajectory of the RL-controlled defect closely follows the prescribed path with little variance, and the absolute error (Fig. \ref{fig:train}b) remains markedly lower compared to the PI controller.
	
	The PI controller, by comparison, again suffers from a delayed response. Its fixed-gain structure does not adjust well to rapid changes in the target trajectory, leading to large error residuals and persistent oscillations. These deficiencies become even more pronounced when the target trajectory is dynamic, as the reactive nature of the PI controller cannot cope with the dynamic non-linear dynamics of the system. Figures~\ref{fig:train}c-d further illustrate this point: while both controllers continuously modulate the activation radius and activity strength, the adjustments of the RL agent are notably more agile and time sensitive. This agility is a direct consequence of the agent’s training, which optimizes the control policy for many episodes and allows it to anticipate the future evolution of the defect motion.
	
	
	To rigorously test the robustness and adaptability of the reinforcement learning (RL) framework, the same trained agent is deployed on three target trajectories that differ substantially from the one used in the training phase and extend the time horizon to $t_f=1000$. In particular, the RL agent's policy is fixed and no further training or fine-tuning is performed, providing a clear measure of its ability to extrapolate from previously learned behavior to new conditions.
	In the first test, the target remains constant over time, $\hat{r}_d(t)=r_0$. Despite having been trained on a dynamic sinusoidal trajectory, the RL agent seamlessly adapts to this simpler static target scenario. Physically, this success highlights the fact that the RL agent has internalized not just the solution to a single trajectory tracking problem but also the mechanisms by which activity modulations drive defect motion. Specifically, the agent learned how the adjustment of the cutoff radius $r_a$ and activity strength $\alpha$ shapes local flow fields, which in turn move the defect radially inward or outward. Using this understanding, the agent quickly stabilizes the defect at the prescribed radial location without oscillations. This result underscores the agent's capacity to handle steady and time-varying goals simply by recognizing that in the static case, fewer corrective actions are needed once the defect is locked in place.
	
	In the second test, the oscillation frequency is halved relative to the training scenario. Halving the frequency implies that the defect must traverse a more gently varying radial path, over a longer timescale. A controller that only memorized specific timings from the training data would struggle under these altered conditions, but Fig.~\ref{fig:test}a shows that the RL agent adapts effectively. From a physics standpoint, a lower frequency trajectory means that the agent has more time between successive turning points of the sine wave. The ability of the RL agent to track this slower oscillation suggests that it has captured the underlying mapping between defect motion and control actions, rather than relying on a rigid schedule of when and how to update $r_a$ and $\alpha$. Observing the state of the defect continuously, especially its instantaneous velocity and deviation from the target radius, the RL policy can adjust the active stress gradient in a measured and consistent way in time. As a result, it remains in synchrony with the new slower oscillatory pattern of the defect.
	
	Lastly, the system is challenged with a composite trajectory $\hat{r}(t)=r+A\sin(\omega t/t_f)+B\cos(\omega t/t_f)$ with $r=12, A=2, B=1, \omega=4\pi$. This test introduces more complex radial variations by superimposing two different oscillatory components. Although the agent was trained on a simpler single-frequency sine wave, it was able to track the composite pattern with high accuracy. The success of the RL agent here is a strong indicator of its generalization to trajectories that it never encountered during training. At a deeper level, this result can be attributed to the underlying policy network of the RL agent, which has effectively learned to respond to deviations between the current and target positions of the defect at each time step. Since the agent continuously senses the current location of the defect, its velocity, and the instantaneous target position (including the target velocity or the rate of change), it can rapidly formulate the necessary adjustments in $r_a$ and $\alpha$. This adaptability ensures that even a more intricate superposition of sinusoidal and cosinusoidal terms does not confound the policy.
	
	
	\textit{Conclusions.}
	In this work, we introduced a reinforcement learning framework for closed-loop control of an active polar defect by dynamically modulating the spatiotemporal activity field. By coupling a continuum hydrodynamic model to an off-policy actor-critic algorithm, the RL agent learns to autonomously discover control protocols that guide the defect along both static and dynamic trajectories. It is found that the RL agent converges rapidly to the desired state, effectively minimizes deviations from target trajectories, and robustly generalizes across diverse scenarios, thereby significantly outperforming standard PI controllers in both static and dynamic settings.

    The fact that a single fixed RL policy can track multiple different target trajectories, ranging from static positions to lower-frequency oscillations, and even composite waveforms, underscores the agent's remarkable ability to generalize. Rather than memorizing a single control law, the RL approach captures the underlying physics governing defect motion in active fluids, enabling seamless adaptation to novel conditions. This shows that reinforcement learning can discover, learn, and exploit the underlying physics of active matter control, as shown by a single policy that generalizes across both static and complex dynamic trajectories. Consequently, these results suggest that RL-based control strategies can serve as robust and versatile platforms for defect manipulation and, by extension, more general applications in self-organizing active materials.
		
	Beyond these immediate demonstrations, the findings highlight the promise of data-driven control strategies for managing the characteristic of emergent and non-linear dynamics of active matter systems \cite{Takatori2025}. The inherent adaptability of the RL-based approach opens avenues for extending control to multidefect environments and designing reconfigurable active materials \cite{Shankar2024}. Moreover, the ability to learn effective policies without requiring complete knowledge of the underlying equations of motion makes this method well suited for real-time experimental implementations in systems such as active fluids, cytoskeletal networks, and other bioinspired active media \cite{Alam2024,Lemma2023}. Future directions will focus on scaling the framework to higher-dimensional control tasks, optimizing energy-efficient actuation protocols, and integrating direct experimental feedback to enable true real-time control. Together, these advances establish reinforcement learning as a powerful tool for engineering and controlling active matter.
	
	\bibliography{Bib}
	
\end{document}


\maketitle
	\section{Supplementary Movie Description}\label{movies}
	Time evolution of the polarity, vorticity, and the spatio-temporal activity field. The simulation environment is fixed with the following set of parameters: $\eta=1, \nu=2, \gamma=0.5, \lambda=15, K_s=5, K_b=7.5,$ and $\chi=-0.5$.
	In each movie, polarity streamlines are shown on the left, colored by Frank free energy density. The background color on the left shows the vorticity field of the flow. The right panel shows the spatiotemporal activity field. Note that the activity field is discretized and hence may appear rotationally asymettric due to the discretization.
	\subsection{Supplementary Movie \rom{1}: Spontaneous rotation of a unit charge defect under fixed spatiotemporal activity}\label{mov1}
The defect moves to a location $r_d$ and rotates at the same radial location (shown as the dashed line in red) for a given fixed activity shown in the right panel.
	\subsection{Supplementary Movie \rom{2}: The reinforcement learning (RL) controlled defect trajectory}\label{mov2}
The target trajectory is set to $\hat{r}(t)=r+A\sin(\omega t/t_f)$ with parameters $r=12, A=2, \omega=4\pi$ and $t_f=500$, shown as a dashed red line on the left at time $t$. The agent is trained on the same moving trajectory and learns to modulate the spatiotemporal activity as shown on the right to keep the defect close to the target trajectory. 
	\subsection{Supplementary Movie \rom{3}: Controlled proportional-integral (PI) defect trajectory}\label{mov3}
The target trajectory is set to $\hat{r}(t)=r+A\sin(\omega t/t_f)$ with parameters $r=12, A=2, \omega=4\pi$, and $t_f=500$, shown as a dashed red line on the left at time $t$. PI controller modulates the spatiotemporal activity as shown on the right but fails to keep the defect sufficiently close to the target trajectory. 
	
	\subsection{Supplementary Movie \rom{4}: The reinforcement learning agent (RL) that controls the defect on an unseen trajectory}\label{mov4}
The target trajectory is set to $\hat{r}(t)=r+A\sin(\omega t/t_f)+B\cos(\omega t/t_f)$ with $r=12, A=2, B=1, \omega=4\pi$, and $t_f=1000$, shown as a dashed red line on the left at time $t$. The same agent as in the SI movie \rom{2} is used to modulate spatio-temporal activity. The agent keeps the defect sufficiently close to the target trajectory. 
	
	\section{Numerical Solution of Active Polar Hydrodynamics}\label{simdetails}
	We solve the following hydrodynamic equations of active polar fluids \cite{kruse_asters_2004,julicher_hydrodynamic_2018},

	\begin{align}
		\frac{\partial \mathbf{p}}{\partial t}+(\mathbf{u}\cdot\nabla)\mathbf{p}+\mathbf{W}\cdot\mathbf{p}=\frac{\mathbf{h}}{\gamma}+\lambda\alpha\mathbf{p}-\nu \mathbf{E}\cdot\mathbf{p}
	\end{align}
	where $\mathbf{p}$ is the polarity vector, $\gamma$ is the rotational viscosity, $\lambda$ is the Onsager coupling constant, and $\nu$ is the flow aligning/tumbling constant. $\mathbf{E}=\frac{1}{2}(\nabla\mathbf{u}+(\nabla\mathbf{u})^\mathrm{T})$ is the strain-rate, and $\mathbf{W}=\frac{1}{2}(\nabla\mathbf{u}-(\nabla\mathbf{u})^\mathrm{T})$  is the spin tensor of the velocity field $\mathbf{u}$. $\mathbf{h}$ is the molecular field obtained via the variational derivative of the Frank free energy density,
	\begin{align}
		\mathcal{F}=\frac{K_s}{2}(\nabla\cdot\mathbf{p})^2+\frac{K_b}{2}(\nabla\times\mathbf{p})^2+\frac{\chi}{2}|\mathbf{p}|^2(1-\frac{1}{2}|\mathbf{p}|^2)
	\end{align}
	with $K_s$ is the material constants of splay, and $K_b$ is the material constant of bend. $\chi$ is a penalty term to enforce the unit magnitude of polarity as a soft-constrained in the system. The force balance in the system is described by
	\begin{align}
		\nabla.(\mathbf{\Sigma^{\mathrm{a}}}+\mathbf{\Sigma^{\mathrm{p}}})=0 && \nabla.\mathbf{u}=0
	\end{align}
	where $\mathbf{\Sigma^{\mathrm{a}}=\alpha(x,t)(\mathbf{p}^2-\frac{1}{2}\mathbf{I}})$ is the active stress driving spontaneous flows and $\mathbf{\Sigma^{\mathrm{p}}}=\mathbf{\Sigma}^s+\mathbf{\Sigma}^{ant}+\mathbf{\Sigma}^{eq}$ is the passive nematic liquid crystal stress comprising of the following symmetric $\mathbf{\Sigma}^s$, anti-symmetric $\mathbf{\Sigma}^{ant}$, and equilibrium stresses $\mathbf{\Sigma}^{eq}$:
	\begin{align}
		\mathbf{\Sigma}^s=2\eta(\nabla\mathbf{u})+\frac{\nu}{2}\Bigl(\mathbf{p}\,\mathbf{h}^T + \mathbf{h}\,\mathbf{p}^T - (\mathbf{p}\cdot\mathbf{h})\,I\Bigr)\\
		\mathbf{\Sigma}^{ant} = \frac{1}{2} \Bigl(\mathbf{p}\,\mathbf{h}^T - \mathbf{h}\,\mathbf{p}^T\Bigr)\\
		\mathbf{\Sigma}^{eq} = -\nabla \mathbf{p}\,\biggl(\frac{\partial \mathcal{F}}{\partial (\nabla \mathbf{p})}\biggr)^{T}.
	\end{align}
	\subsection{Non-dimensionalization}\label{nd}
	
	We define \emph{active length scale} $L_a = \sqrt{\frac{\delta K}{\alpha}}$ and \emph{active time scale} $ \tau_a = \frac{1}{\alpha}$, where $\delta K=K_b-K_s$.
	Using these, one can introduce the dimensionless variables
	\begin{align}
		\hat{\mathbf{x}} = \frac{\mathbf{x}}{L_a}, 
		\quad
		\hat{t} = \frac{t}{\tau_a} = \alpha t, 
		\quad
		\hat{\mathbf{u}} = \frac{\mathbf{u}}{U_a},
	\end{align}
	where $U_a = \frac{L_a}{\tau_a} = \sqrt{\alpha K}.$
	We choose the parameters,  $\eta=1, \nu=2, \gamma=0.5, \lambda=15, K_s=5, K_b=7.5,$ and $\chi=-0.5$ for the simulation environment. The simulation is that the disk is of radius $R=25$. $r_a, \alpha$ is chosen by the agent within a range of $[0,R], [-0.2,0]$

	\subsection{Numerical simulation}
	
	The polarity field is discretized on a set of unformally randomly distributed points within the simulation domain, a disk of radius $R=25$ with an average spacing of 1.5 units. The polarity is initialized with 
	\begin{align}
		\mathbf{p}(x,y)=\begin{bmatrix}
			p_x\\
			p_y
		\end{bmatrix}
			=
		\begin{bmatrix}
				\cos(\phi)*x-\sin(\phi)*y\\
				\sin(\phi)*x+\cos(\phi)*y
		\end{bmatrix}		
	\end{align}	
	and $\phi=0.1\pi$. The boaundry is fixed with the intial condition as Dirichlet boundary conditions for the polarity. Differential operators are computed using Discretization-Corrected particle strength exchange \cite{schrader_discretization_2010,singh_c_2021}.  The time evolution of the polarity is computed using an adaptive Adams-Bashforth-Moulton predictor-corrector time integration with an initial time step of $0.1$ as described previously in \cite{Singh2023}. The time evolution continues until the final time $t_f$. 
	
	The velocity field is computed at every time step by iteratively correcting the pressure and solving the implicit system of incompressible force balance equations with no-slip boundary conditions for the flow. At each time step, the resulting linear system of equations is numerically solved using the iterative GMRES solver implemented in the PETSc software library~\cite{balay_efficient_1997}. We checked that using a higher resolution yields the same results, confirming convergence in space and time. The convergence of the algorithms used has also been demonstrated previously \cite{Singh2023}.
	The entire simulation environment is implemented using a custom C++ expression system, within the OpenFPM library \cite{incardona_openfpm_2019}. 
	
	\section{Reinforcement Learning with VRACER}\label{rldetails}
	
	In our work, we employ an off-policy actor-critic algorithm VRACER \cite{pmlr-v97-novati19a,novati_controlled_2019} as implemented in the Korali framework \cite{Martin2022} to learn robust control policies for active polar defect dynamics. In the following, we describe the key components and hyperparameters of our RL setup.
	
	\subsection{Environment and MPI Distribution}
	The simulation environment is encapsulated in a custom function that interfaces with our simulation model environment. This environment:
	\begin{itemize}
		\item Initializes the simulation.
		\item Retrieves the current state vector and broadcasts actions across MPI workers.
		\item Advances the simulation by applying the action, computes the reward based on the deviation from a target trajectory, and updates the state.
		\item Terminates the episode when a predefined maximum number of steps is reached.
	\end{itemize}
	Distributed execution is managed through MPI; the experiment is configured with a user-defined number of cores to ensure efficient parallelization.
	
	\subsection{State and Action Spaces}
	The state vector comprises six variables:
	\begin{enumerate}
		\item \textbf{Defect Radial Position} ($r_d$) : The radial location of the defect in polar coordinates.
		\item \textbf{Activity Radius} ($r_a$): The activation region of the active stress.
		\item \textbf{Defect Speed} ($v_d$): $\frac{r_d(t)-r_d(t-\delta t)}{\delta t}$, instantaneous speed based on the previous location of the defect.
		\item \textbf{Activity Strength} ($\alpha$): Strength of the activity field.
		\item \textbf{Moving Target} ($\hat{r}_t$): The location of the target defect at time $t$.
		\item \textbf{Moving Target Derivative} ($\dot{\hat{r}}_t$): Derivative of the target trajectory at time $t$.
	\end{enumerate}
	The action space is two-dimensional, controlling:
	\begin{itemize}
		\item \textbf{Activity Radius} ($r_a$): bounded between $0$ and $25$, with an initial exploration noise of $3.5$.
		\item \textbf{Activity Strength} ($\alpha$): bounded between $-0.2$ and $0.0$, with an initial exploration noise of $0.02$.
	\end{itemize}
	
	\subsection{Algorithm Configuration and Hyperparameters}
	The RL problem is formulated as a continuous control task using VRACER with the following parameters:
	\begin{itemize}
		\item \textbf{Episodes Per Generation:} $1000$
		\item \textbf{Termination Criteria (Max Generations):} $1$
		\item \textbf{Discount Factor:} $0.995$
		\item \textbf{Learning Rate:} $0.01$
		\item \textbf{Mini Batch Size:} $100$
		\item \textbf{Experience Replay:} Enabled with a maximum size of $100\,000$ and a starting size of $1$. Furthermore, the off-policy importance sampling parameter $\beta$ is set to $0.3$.
	\end{itemize}
	
	\subsection{Neural Network Architecture}
	The policy and value functions are approximated by a neural network, which is configured as follows:
	\begin{itemize}
		\item \textbf{Hidden Layers:} The network comprises a linear layer with $16$ output units followed by an element-wise Tanh activation function.
		\item \textbf{Policy Distribution:} A Clipped Normal distribution is used to represent the policy.
		\item \textbf{Optimizer:} The Adam optimizer is used to update the network weights.
	\end{itemize}
	Rescaling of the state and reward is applied to improve the stability of the training.
	
	\subsection{Algorithm Execution}
	During training, the VRACER algorithm interacts with the defect simulation environment to collect samples. Each sample includes the current state, the action chosen by the policy, the resulting reward, and the next state. After every action step, the collected experiences are stored in an experience replay buffer. Mini-batches of experiences are then drawn to update the neural network weights, thus refining both the policy and the value function.
	
	The complete experiment is executed by invoking the Korali engine with the specified configuration. This setup facilitates efficient exploration of the state-action space and robust learning of control policies for the active polar fluid system.

	\subsection{PI Controller}\label{pidetails}
	
	As a baseline for comparison with our RL-based approach, we also implement a classical proportional-integral (PI) controller to generate control actions based on the state of the defect. The PI controller adjusts two control signals: the activity radius and the activity strength. The tracking error is defined as
	\begin{align}
		e(t) = \hat{r}(t) - r_d(t),
	\end{align}
	where \(\hat{r}(t)\) is the target radial position at time $t$ and \(r_d(t)\) is the current defect radial position.
	
	The control law for the activity radius is given by
	\begin{align}
		r_a(t) = K_p\,e(t) + K_i \int_{0}^{t} e(\tau) \, d\tau,
	\end{align}
	with the proportional gain \(K_p = 2\times10^{-3}\) and the integral gain \(K_i = 11\times10^{-3}\).
	
	Similarly, the control signal for the activity strength is computed as
	\begin{align}
		\alpha(t) = K_{p_a}\,e(t) + K_{i_a} \int_{0}^{t} e(\tau) \, d\tau,
	\end{align}
	with \(K_{p_a} = -1\times10^{-5}\) and \(K_{i_a} = -9\times10^{-5}\). To ensure that the strength of the activity remains within its physical limits, the output \(\alpha(t)\) is cut to be between \(-0.2\) and \(0.0\).	

	\subsection{Static Target Training}\label{starget}
		\begin{figure}
			\centering
			\setlength{\tabcolsep}{10pt}
			\captionsetup[subfloat]{captionskip=0pt} 
 			\subfloat[]{ \includegraphics[scale=0.35]{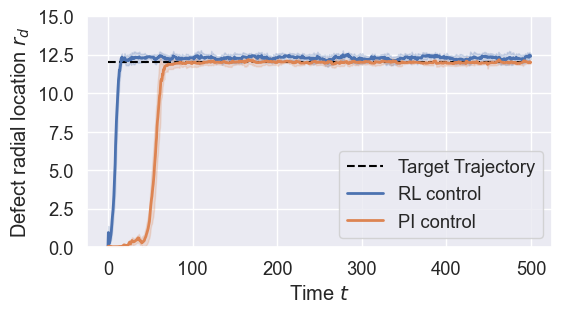}} 
			\subfloat[]{\includegraphics[scale=0.35]{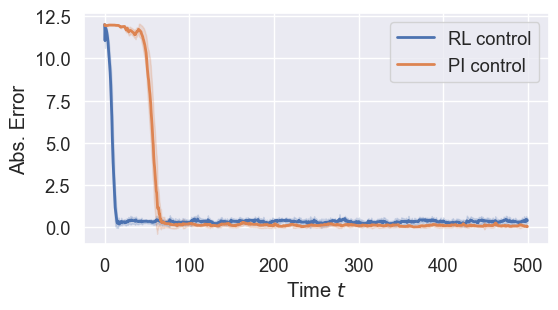}} 	\subfloat[]{\includegraphics[scale=0.35]{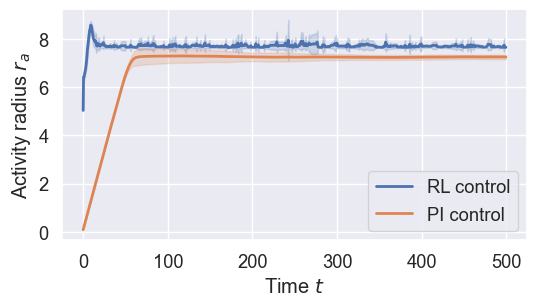}}
			\vspace*{-0.25cm}     
			\caption{ 
				Comparison of RL vs PI control when the RL agent is trained on the static target trajectory. Mean and standard deviations over 10 trajectories are shown. (a) Time evolution of the defect’s radial position $r_d$. The dashed line indicates the target trajectory, while the solid lines show the RL-controlled (blue) and PI-controlled (orange) defect positions. (b) Absolute tracking error for the RL (blue) and PI (orange) controllers. The RL approach achieves smaller deviations over time when compared to PI. (c) Time evolution of the activity radius $r_a$ as determined by each controller. Note that the activity is fixed in this case.}
			\label{fig:static}      
		\end{figure}
		
\begin{figure}
		\centering
		\setlength{\tabcolsep}{10pt}
		\captionsetup[subfloat]{captionskip=0pt} 
		\subfloat[]{ \includegraphics[scale=0.35]{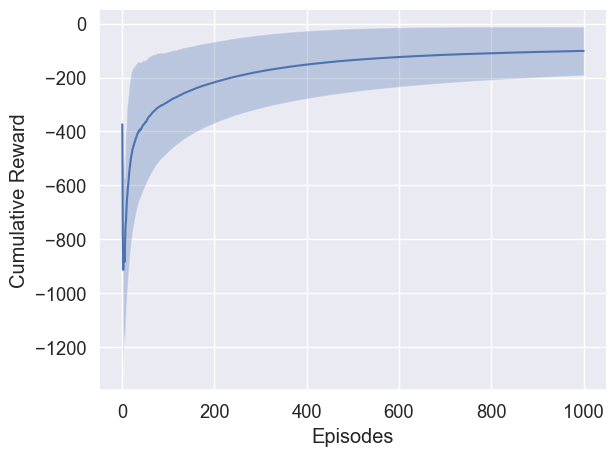}} 
		\subfloat[]{\includegraphics[scale=0.35]{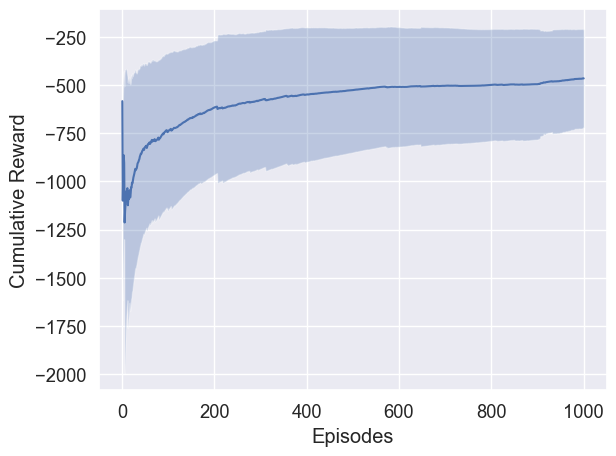}} 
		\vspace*{-0.25cm}     
		\caption{ 
			Cumulative reward obtained by the agent during training. (a)
			Cumulative reward obtained when training on the static trajectory $\hat{r}(t)=r_0=12$ with only controlling $r_a$ and keeping $\alpha=0.11$. 
			(b) Cumulative reward obtained when training on the dynamic trajectory \(\hat{r}(t)=r+A\sin(\omega t/t_f)\) with \(r=12\), \(A=2\), \(\omega=4\pi\), and \(t_f=500\) controlling both $(r_a, \alpha)$.}
		\label{fig:cumreward}      
	\end{figure}

	In this set of experiments, the RL agent is trained to maintain the defect at a fixed radial target, \(\hat{r}_d(t)=r_0\) (with \(r_0=12\)). During each episode, the agent receives as input the current position, speed, and state of the active field of the defect, and then outputs control actions. Specifically, adjustments to the activity radius \(r_a\) and the activity strength \(\alpha\). The instantaneous reward is defined as the absolute deviation between the defect position and the target position, and the cumulative reward is recorded over each episode. As shown in Fig.~S\ref{fig:cumreward}a, the cumulative reward rapidly converges with the training, indicating that the agent quickly learns to reduce positional error. Concurrently, Fig.~S\ref{fig:static} compares the time evolution of the defect radial position under RL control against that achieved with a conventional PI controller. The RL agent maintains the defect near the prescribed target with minimal oscillations, whereas the PI controller exhibits persistent transient oscillations and slower convergence to the target. These results clearly demonstrate the superior performance and learning efficiency of the RL approach for static target control.
	
	\subsection{Dynamic Target Training }\label{dtarget}	
	For dynamic target training, the target trajectory is specified as \(\hat{r}(t)=r+A\sin(\omega t/t_f)\) with \(r=12\), \(A=2\), \(\omega=4\pi\), and \(t_f=500\). In this scenario, the RL agent is tasked with tracking a continuously varying target, which poses a greater challenge due to the time-dependent nature of the control problem. At each time step, the agent observes the current state of the defect, including its radial position, speed, and instantaneous target position and derivative, and accordingly modulates the control parameters \(r_a\) and \(\alpha\) to minimize tracking error. Figure~S\ref{fig:cumreward}b shows the evolution of cumulative reward in training episodes in the dynamic regime, evidencing rapid convergence as the agent refines its control policy. The figure in the main text presents a side-by-side comparison between the RL-controlled and PI-controlled defect trajectories. The RL agent consistently follows the sinusoidal target with minimal deviation, while the PI controller suffers from larger tracking errors and pronounced oscillations. These observations underscore the effectiveness of the RL framework in dynamically adjusting to a time-varying target, thereby outperforming traditional control strategies in managing the complex dynamics of active polar defects.
	
		\subsection{Inference}	
	During inference, the performance of the RL agent is enhanced by refining the estimation of the location of the defect by interpolation using a paraboloid kernel. In the training phase, the defect position was determined solely by identifying the minimum value on a discrete point location. However, for inference, we apply the paraboloid kernel to interpolate between grid points, thereby generating a smoother and more accurate estimate of the defect's position. This approach effectively reduces the discretization error inherent in the simulation and allows the agent to more accurately track the defect in a continuously evolving environment. The improved resolution obtained through this interpolation not only enhances the control performance but also demonstrates the generalizability of the RL agent to continuous state spaces, despite its training on a coarser, discrete representation. The inference performance on various trajectories is shown in the main text.
		
	\bibliographystyle{ieeetr}
	\bibliography{Bib}